\documentclass[         %
aps,                    
prd,                    
showpacs,               
superscriptaddress,    
nofootinbib,            
showkeys,               %
preprintnumbers,        %
floatfix]               
{revtex4}               
\usepackage{graphicx,longtable,amsmath}
\usepackage{enumerate}

\begin{document}
\title{Neutrino oscillations above black hole accretion disks: disks with electron-flavor emission
}
\author{A.~Malkus}
\email{acmalkus@ncsu.edu}
\author{J.~P.~Kneller}
\email{jim_kneller@ncsu.edu}
\affiliation{Department of Physics, North Carolina State University, Raleigh, NC 27695 USA.}
\author{G.~C.~McLaughlin}
\email{gcmclaug@ncsu.edu}
\affiliation{Department of Physics, North Carolina State University, Raleigh, NC 27695 USA.}
\author{R.~Surman}
\email{surmanr@union.edu }
\affiliation{Department of Physics and Astronomy, Union College, Schenectady, NY 12308 USA}

\date{\today}
\begin{abstract}
Black hole accretion disks can form through the collapse of rotating massive stars. 
These disks produce large numbers of neutrinos and antineutrinos of electron flavor that can influence energetics and nucleosynthesis. Neutrinos are produced in sufficient numbers that, after they
are emitted, they can undergo flavor transformation facilitated by the neutrino self interaction.
We show that some of the neutrino flavor transformation phenomenology for accretion disks is similar to  that of the supernova case, but also, we find the disk geometry lends itself to different transformation behaviors.
These transformations strongly influence the nucleosynthetic outcome of disk winds.
\end{abstract}
\medskip
\preprint{}
\maketitle

\vskip 1.3cm

\section{Introduction}
\label{Section: Introduction}
Accretion disks are a compelling astrophysical setting, the properties of which are still being understood.
Sophisticated numerical models have examined the origins of the disks and their structure.
The disks may arise out of mergers between neutron stars \cite{Eichler:1989ve, Ruffert:2001gf}, 
between a neutron star and a black hole \cite{Narayan:1992iy}, 
or as a result of some stellar collapses \cite{Paczynski:1997yg, MacFadyen:1998vz, MacFadyen:1999mk}.
It has been suggested that accretion disks could play an important role in the production of gamma ray bursts \cite{Ruffert:1998qg, Meszaros:2001vi} and they
have also been studied as possible sites for nucleosynthesis 
in the absence of neutrino oscillations \cite{Pruet:2002ky, Pruet:2003yn, Surman:2003qt, Surman:2008qf, Kizivat:2010ea, Metzger:2010sy,Caballero:2011dw,  Wanajo:2011vy}.
Disks that originate from stellar collapse with sufficiently high accretion rates are understood to have a region of trapped neutrinos with mean energies of tens of MeV.
These disks emit primarily electron and anti-electron type neutrinos and relatively little mu and tau type.
The emission surface of the neutrinos generally exceeds that of the antineutrinos, but the antineutrinos have higher temperatures
\cite{Matteo:2002ck, Surman:2003qt, Chen:2006rra}.
We focus on these features of the stellar collapse case.

Beginning in the early 1990s, it was realized \cite{Pantaleone:1992eq} that coherent forward scattering of neutrinos could impact neutrino oscillation, 
resulting in coherence between neutrinos of different energies and parametric resonances \cite{Samuel:1993uw}.
Therefore, we expect that neutrino-neutrino interactions will play an important role in neutrino oscillations above disks.
Neutrino oscillations involving neutrino-neutrino interactions have been extensively studied in general and in the contexts of 
the early universe and supernovae \cite{Balantekin:2006tg, 
Barbieri:1990vx,
Qian:1995ua, 
Fogli:2007bk,
Gava:2009pj, Galais:2009wi, Duan:2007fw, 
Duan:2010bf, Duan:2010bg,Duan:2009cd, EstebanPretel:2007bz, Duan:2006an, 
Fuller:2005ae,Banerjee:2011fj,
Sawyer:2008zs}.
When restricted to two flavors, the numerical results in regions of high neutrino density have been analyzed 
in terms of an analogy to precession and nutation of spins in a magnetic field \cite{Pastor:2001iu,Duan:2006jv,Hannestad:2006nj}.
This analogy is called the Neutrino Flavor Isospin (NFIS) picture.
Another approach was to use the techniques of BCS theory \cite{Pehlivan:2011hp}. 
The early calculations assumed that interacting neutrinos shared a common history, in what is known as the single angle approximation \cite{Qian:1994wh}.
Calculations that compute the histories of neutrinos across many trajectories resulted in effects not seen in the single angle approximation, 
including the decoherence of different energy modes \cite{Duan:2005cp,EstebanPretel:2007ec}.
Numerical calculations \cite{Dasgupta:2009mg, Dasgupta:2010cd, Fogli:2009rd, Duan:2008za}
showed that oscillations split the neutrino spectra among flavors at discrete energies.
These so-called spectral splits depend on the adiabaticity of the oscillations \cite{Raffelt:2007xt,Pehlivan:2011hp}.
They can be understood in terms of the NFIS picture as a magnetic resonance phenomenon \cite{Galais:2011gh}.

Neutrino-neutrino interactions above disks were studied by Dasgupta et al. \cite{Dasgupta:2008cu}.
They generalized self-maintained coherence behavior --where neutrinos of different energies oscillate as a single mode-- to a disk geometry.
For their disk geometry, Dasgupta et al. studied two identical, circular disks, without black holes at the center, emitting both neutrinos and antineutrinos, 
with number densities fixed to a constant ratio, so that neutrinos dominated the self-interaction term.
Coherent forward scattering of neutrinos on electrons (matter) was included 
by assuming a small mixing angle.
With this model, a single ``nutation'' region, similar to that found in supernovae, was found when the neutrino densities became sufficiently low.

In this paper, we create disk models that reflect the qualitative understanding we have of black hole accretion disks that originate from stellar collapse and include an explicit matter interaction term.
We choose our neutrino and antineutrino disks to have different temperatures and different radii, so that the ratio between the neutrino flux densities will not be fixed, but will vary with position.

Typically, the antineutrino disk will be hotter than the neutrino disk, but smaller. 
Therefore, there are two basic relationships the densities may have above the disks. 
Neutrinos may always dominate or antineutrinos dominate near the disk while neutrinos dominate further away. 
Our disks have ``holes'' in the center, from which no neutrinos or antineutrinos are emitted. 
This region corresponds to the space close to the black hole within its last stable orbit.
To these disks, we add a nontrivial electron density.

We also consider the role that neutrino flavor transformation plays in nucleosynthesis. 
The impact of matter-enhanced neutrino flavor transformation core collapse supernovae nucleosynthesis has been considered in e.g. \cite{Yoshida:2006qz, Yoshida:2006sk, Qian:1993dg, Fuller:1998kb} and 
the impact of self interaction on supernova nucleosynthesis has been considered in \cite{Duan:2010af, Chakraborty:2009ej}.
However, the impact of flavor transformation in disk nucleosynthesis has not been previously evaluated.

In section \ref{sec:disk-model} we outline our two general disk models and in section \ref{neutrinos} we outline our method. 
In section \ref{calculation}, we perform a single angle numerical computation of three flavor neutrino oscillations in these disks.
In section \ref{analysis}, we analyze the results, and  find a new type of transition region in accretion disks.
In section \ref{nucleosynthesis} we discuss the ramifications for nucleosynthesis and in section \ref{conclusions} we conclude.

\section{Disk Model \label{sec:disk-model}}

\subsection{General Features of Accretion Disks}\label{qualitativeFeatures}
Our goal is to use a model that reproduces some general features of disks. 
The material near the plane of the disk is hot and protons and neutrons exist as free nucleons participating in the reactions of equation \ref{neutrinoCaptureEmission} as well as other scattering interactions.

\begin{equation}
  \begin{aligned}
    n + \nu_e   \rightleftharpoons & e^- + p\\
    p + \bar\nu_e \rightleftharpoons & e^+ + n\label{neutrinoCaptureEmission}
  \end{aligned}
\end{equation}
For some disks with low accretion rates, below $\sim 0.1$ $M_\odot/s$, the reactions primarily proceed from right to left 
since electron neutrinos and electron antineutrinos are not trapped \cite{Popham:1998ab}. 

As the accretion rate increases above $ \sim 0.1 \, M_\odot/s$, the rates for neutrino interactions increase as well.
Where the disk is hottest, near the plane of the disk, the neutrinos may be trapped \cite{Matteo:2002ck}.
Further away, they can stream freely.
For disks with high accretion rates, $\sim 1 \, M_\odot/s$, both electron neutrinos and antineutrinos are trapped \cite{Surman:2003qt, Chen:2006rra}, although the emission surface of the electron neutrinos is more extended along the plane of the disk.
In addition, electron antineutrinos begin to free stream closer to the plane of the disk than electron neutrinos.
Therefore, electron antineutrinos are hotter than the electron neutrinos when they are both free streaming, but the electron neutrinos originate from a larger emission surface.

Other flavors of neutrinos, including $\nu_\mu$ and $\nu_\tau$ are trapped in disks with high accretion rates, eg $\sim 10$ $M_\odot/s$ \cite{McLaughlin:2006yy, Ruffert:2001gf}. They can produce a range of nucleosynthesis \cite{Surman:2005kf} and are interesting because they may be sites for $r$-process nucleosynthesis. We will not address them here, because they do not typically originate from stellar collapse.

Given these considerations, we focus on two disks that describe two qualitatively different pictures.  
The flux above a disk can be antineutrino dominated close to the emission surface (since the antineutrinos are hotter) 
but neutrino dominated far from the disk since the neutrino emission surface is larger, and the disk is radiating net lepton number.
This is the situation that one would expect from the type of steady
state disk models described in Surman and McLaughlin \cite{Surman:2003qt} and Chen and Beloborodov in \cite{Chen:2006rra}. 
The other situation that can occur is that the flux above the disk is always neutrino dominated.

\subsection{Our Models}\label{models}
We use two models that reproduce the 
features discussed in the section \ref{qualitativeFeatures}.
We take disks to be thin, geometrically flat with a hole in the center with a constant temperature throughout.
These temperatures determine the neutrino fluxes entirely since we characterize the spectrum with 
Fermi-Dirac distributions with zero chemical potential.
The disks emit only electron neutrinos and antineutrinos and we assume that the neutrinos and antineutrinos emitted from the disks follow straight line paths, not relativistic paths.

We chose 
two relationships between the neutrino and antineutrino disks.
These disks are summarized in table \ref{whatAreOurModels}.
First, we look at a disk where the neutrinos are slightly cooler than antineutrinos but have a larger disk.
We call this model A. 
It is the most qualitatively similar to the proto-neutron star: the trajectory we follow above this disks will be dominated by neutrinos. 
We also look at a disk which is motivated by the disks discussed in \cite{Surman:2003qt, Chen:2006rra} where antineutrinos are significantly hotter than the neutrinos, but again the 
neutrinos have a larger disk.
We call this model B.
The trajectory above this disk will start out dominated by antineutrinos near the disk surface and at some further position will be dominated by neutrinos. 

We take the electron densities to be consistent with a wind type of outflow as discussed e.g. in \cite{Surman:2005kf}, 
and use the same electron density in both models. We give the explicit form of the parametrization in section \ref{nucleosynthesis}.

We examine both disks in the normal and inverted hierarchies.
The vacuum neutrino mixing parameters were taken to be the mass squared differences, 
$m_2^2-m_1^2=\delta m_{21}^2=7.59\times10^{-5}$ eV$^2$, $\left|m_3^2-m_2^2\right|=\left|\delta m_{32}^2\right|=2.43\times10^{-3}$ eV$^2$; 
and the mixing angles, $\theta_{13}=9^\circ$, $\theta_{12}=34.4^\circ$ and $\theta_{32}=45^\circ$.
These choices are consistent with the current PDG values \cite{pdg}.

\begin{table}
\begin{tabular}{|l|l|l|l|l|l|}
\hline
Model & $T_\nu$ & $T_{\bar{\nu}}$ &$R_\nu$ & $R_{\bar{\nu}}$  & $R_0$\\\hline
A     & 3.2 MeV & 3.4 MeV & $1.5\times 10^7$ cm & $10^7$ cm& $3.2\times10^6$ cm\\\hline
B     & 3.2 MeV & 4.1 MeV & $1.5\times 10^7$ cm & $10^7$ cm& $3.2\times10^6$ cm \\
\hline
\end{tabular}\caption{Parameters corresponding to models in this paper, 
including $T_\nu$, the temperature of the neutrino emitting surface; $T_{\bar{\nu}}$, the temperature of the antineutrino emitting surface;
$R_\nu$, the outer radius of the neutrino emitting disk; $R_{\bar{\nu}}$, the outer radius of the antineutrino emitting disk; 
and $R_0$, the inner radius of both the neutrino and antineutrino emitting disks.\label{whatAreOurModels}}
\end{table}

\section{Neutrinos}\label{neutrinos}
Neutrino oscillations are governed by an equation with the ambient neutrino potential, 
the vacuum contribution and a contribution from interactions with electrons and positrons, 
\begin{equation}
  \begin{aligned}
    i\frac{d}{d t} S(\Omega,q,\mathbf{x},t) =\left(H_V(q)+H_e(\mathbf{x},t)+H_{\nu\nu}(\mathbf{x},t)\right) S(\Omega,q,\mathbf{x},t)\label{schroedinger}
  \end{aligned}
\end{equation}
where $t$ is the time elapsed, $\mathbf{x}$ is the neutrino position, and
  \begin{equation}
    \begin{aligned}
H_{\nu\nu}(\mathbf{x},t)=\sqrt{2}G_F\int_0^\infty  dq\int_\Omega \left(1-\cos\Theta_{qq'}\right)\left(\rho_\nu(\Omega,q,\mathbf{x},t) dn_\nu(\Omega,q,\mathbf{x},t)-\rho_{\bar{\nu}}(\Omega,q,\mathbf{x},t)dn_{\bar{\nu}}(\Omega,q,\mathbf{x},t)\right)
  \end{aligned}
\end{equation}
The integration variable, $q$ is the energy of an ambient neutrino that shapes the potential. 
The infinitesimals, $dn_\nu(\Omega,q,\mathbf{x},t)$ and $dn_{\bar{\nu}}(\Omega,q,\mathbf{x},t)$ are the densities of neutrinos and antineutrinos and $\Theta_{qq'}$ is the angle between a neutrino with momentum $q'$ and an ambient neutrino with momentum $q$.
The information about density evolution is contained in $S(\Omega,q,\mathbf{x},t)$, the scattering matrix that evolves the neutrino density matrix, $\rho_\nu(\Omega,q,\mathbf{x},t)$: $\rho_\nu(\Omega,q,\mathbf{x},t)=S(\Omega,q,\mathbf{x},t) \rho_\nu(\Omega,q,\mathbf{x},0)S^\dagger(\Omega,q,\mathbf{x},t)$.
The antineutrino density matrix, $\rho_{\bar{\nu}}(\Omega,q,\mathbf{x},t)$ is similarly evolved by another scattering matrix, $\bar{S}(\Omega,q,\mathbf{x},t)$.
We will take both the neutrino and antineutrino density matrices to start with all of the neutrinos and antineutrinos in the electron flavor.
The matrix, $H_V$ is the vacuum Hamiltonian. 

The matter Hamiltonian is proportional to $V_e(\mathbf{x},t)=\sqrt{2}G_F(N_{e^-}(\mathbf{x},t)-N_{e^+}(\mathbf{x},t))$, the contribution from neutrinos interacting with electrons and positrons, where
$N_{e^\mp}(\mathbf{x},t)$ is the electron(positron) density which is then scaled by $\sqrt{2}G_F$, where $G_F$ is Fermi's constant.

\begin{figure}[t]
\begin{center}
\includegraphics[scale=0.3]{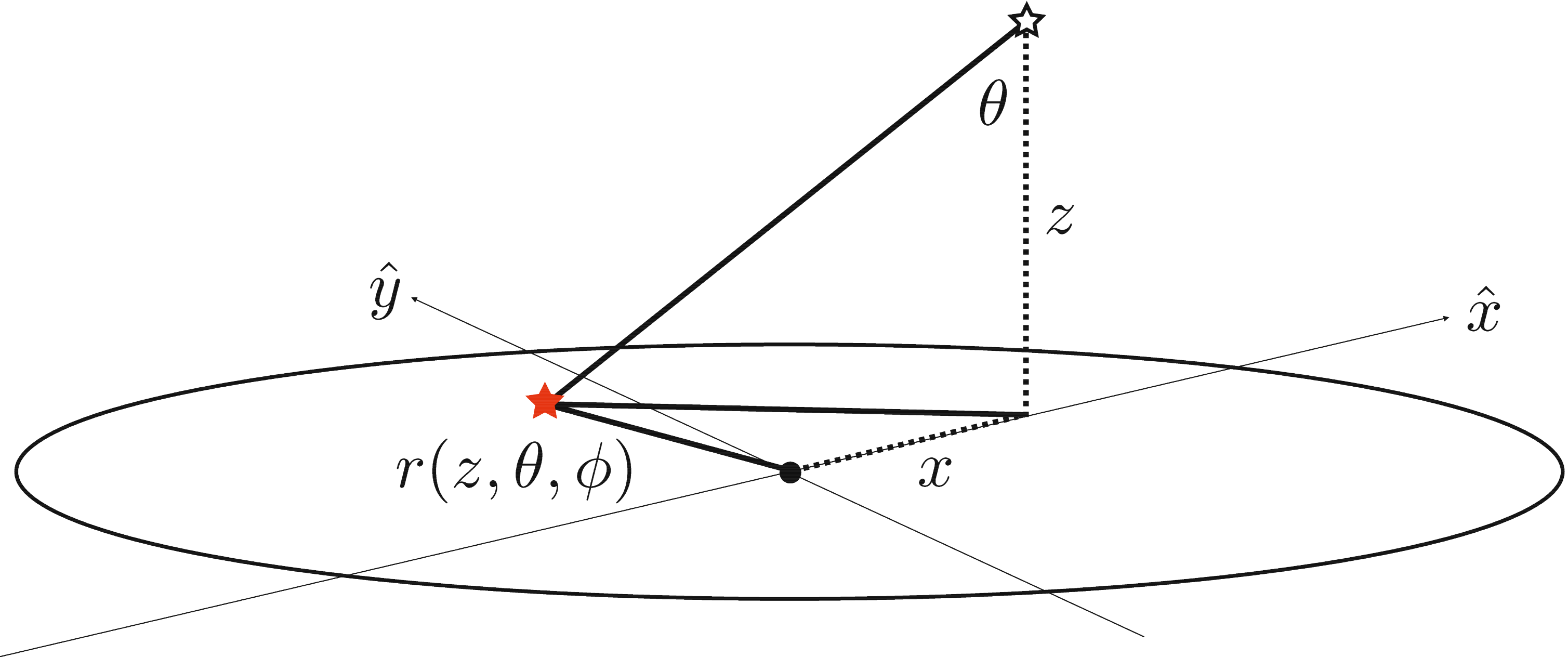}
\caption{A side view of the disk.
The solid (red) star indicates the origin of the neutrino on a flat disk with a straight-line trajectory.
The empty star indicates the current position of the neutrino above the disk.}
\label{defineTheta}
\end{center}
\end{figure}

\begin{figure}[t]
\begin{center}
\includegraphics[scale=0.45]{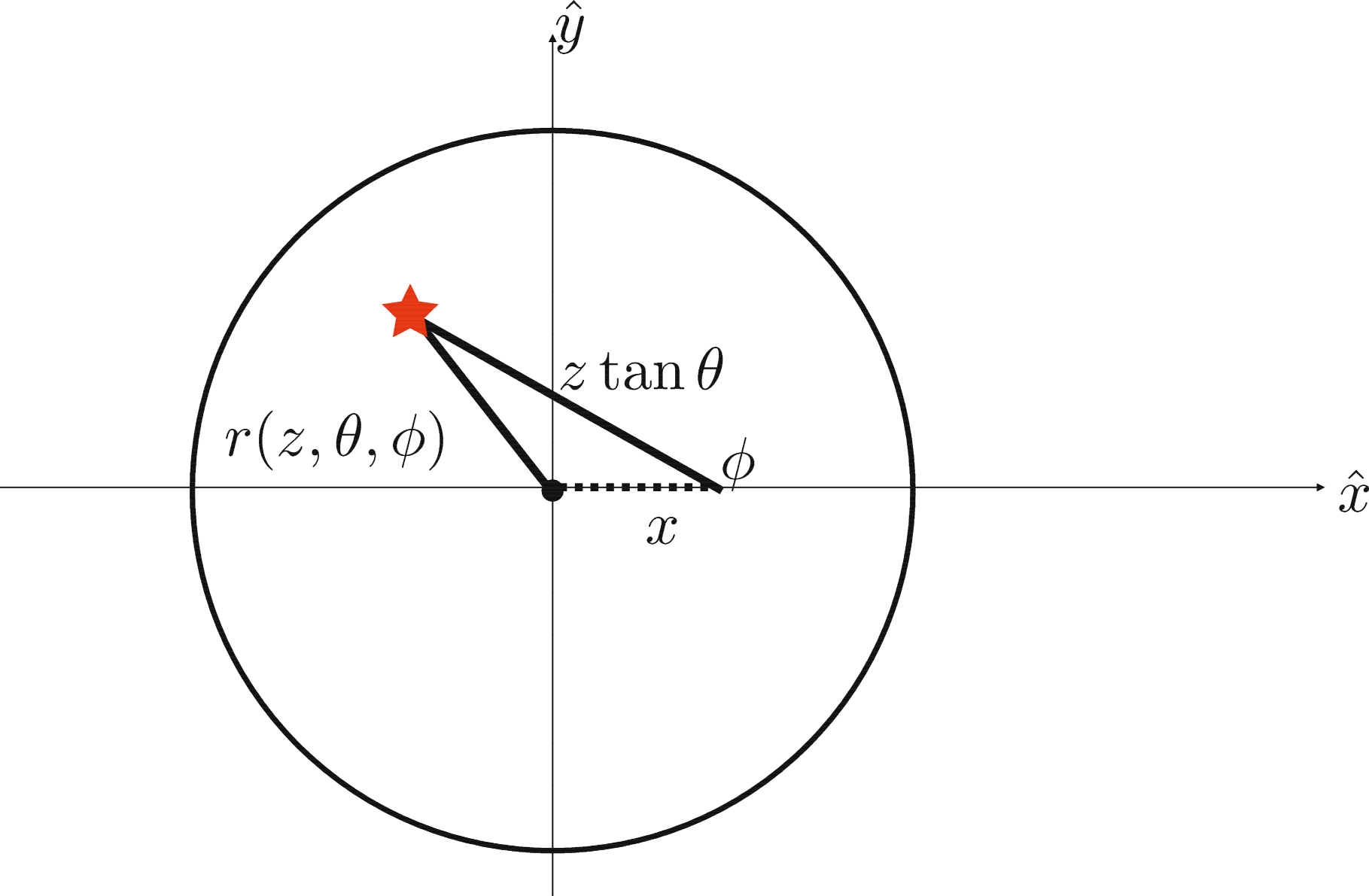}
\caption{A bird's eye view of the disk. The $z$ axis is coming out of the page.
The solid (red) star indicates the origin of the neutrino on the disk.
\label{definePhi}}
\end{center}
\end{figure}

The difference between the accretion disk case and the protoneutron star is the geometry of the source.
To handle the symmetries of the disk, we define a coordinate system based on the location of a neutrino, as shown in Figs. \ref{defineTheta} and \ref{definePhi}. 
We consider a neutrino that is a distance $x$ from the center of the disk and a distance $z$ above. 
From these coordinates, we define $\theta$, the angle between the $z$ axis and the neutrino momentum.
We also define $\phi$, the angle between the $x$ axis and the projection of the neutrino momentum.
This information completely characterizes straight-line trajectories.

In the coordinate system of Figs. \ref{defineTheta} and \ref{definePhi}, the angle between the ambient neutrino with coordinates $\phi$ and $\theta$, 
and the other neutrino with coordinates $\phi'$ and $\theta'$ can be written
\begin{equation}
  \cos\Theta_{qq'}=\cos\theta'\cos\theta+\sin\theta'\sin\theta\cos(\phi'-\phi),
\end{equation}
for all pairs of neutrino trajectories.
The electron neutrino and antineutrino densities are $dn_\nu(\Omega,q,\mathbf{x},t)$ and $dn_{\bar{\nu}}(\Omega,q,\mathbf{x},t)$ respectively.

\begin{equation}
  dn_{\nu(\bar{\nu})}(\Omega,q,\mathbf x,t)=\frac{\Phi_{\nu(\bar{\nu})}(q)d\phi d(\cos\theta)}{4\pi},
\end{equation}
where $\Phi_{\nu(\bar{\nu})}$ is the flux of neutrinos(antineutrinos).

In this work, we use the single angle approximation. 
This assumption factorizes the angular and energy integrals in equation \ref{schroedinger} 
and is equivalent to assuming all neutrinos along every trajectory behave identically.
Because we apply a single angle approximation and because we take the disks to have constant temperature, we can factor the integral of equation \ref{schroedinger} into more manageable parts,
\begin{equation}
  \begin{aligned}
   \frac{H_{\nu\nu}(\mathbf x, t)}{\sqrt{2}G_F}=&\int_0^\infty  dq\int_\Omega \left(1-\cos\Theta_{qq'}\right)\left(\rho_\nu(\Omega,q,\mathbf x,t) dn_\nu(\Omega,q,\mathbf x,t)-\rho_{\bar{\nu}}(\Omega,q,\mathbf x,t)dn_{\bar{\nu}}(\Omega,q,\mathbf x,t)\right)\\
   =&\frac{1}{4\pi}\int_0^\infty \Phi_{\nu}(q) dq\int_{\Omega_\nu} \left(1-\cos\Theta_{qq'}\right)\rho_\nu(\Omega,q,\mathbf x,t)d\phi d(\cos\theta)\\
   &-\frac{1}{4\pi}\int_0^\infty \Phi_{\bar{\nu}}(q) dq\int_{\Omega_{\bar{\nu}}} \left(1-\cos\Theta_{qq'}\right)\rho_{\bar{\nu}}(\Omega,q,\mathbf x,t)d\phi d(\cos\theta).
  \end{aligned}
\end{equation}
where the $\Omega_{\nu(\bar{\nu})}$ reminds us to take the integral over the appropriate range of angles for the neutrino (antineutrino) disk.
This term can be further simplified and the $\theta$ integrals can be performed analytically.
\begin{equation}
  \begin{aligned}
 H_{\nu\nu}(\mathbf x,t)=&\frac{-G_F z}{\sqrt{2}\pi }\int_0^\infty\left(\Phi_\nu(E)\rho_\nu(E,\mathbf x,t)\int_{R_0}^{R_\nu}C(r, \mathbf x,t)rdr-\Phi_{\bar\nu}(E)\rho^*_{\bar\nu}(E,\mathbf x,t)\int_{R_0}^{R_{\bar{\nu}}}C(r, \mathbf x,t)rdr\right) dE,
  \end{aligned}
\end{equation}
where
\begin{equation}
  C(r, \mathbf x,t)=\frac{ \pi   \left((l+m)\left(z\cos\theta' + x\sin\theta'\cos\phi'\right)  - 4x\sin\theta' \cos\phi'\left(x^2+z^2\right)\right)}{2\left(lm\right)^{3/2}}-\frac{2  E\left(\frac{m-l}{m}\right)}{\sqrt{m} l},
\end{equation}
with $R_0$ is the last stable orbit, $R_{\nu(\bar{\nu})}$ is the radius of the (anti)neutrino disk, $E$ is the (anti)neutrino energy $l=(x-r)^2+z^2$, $m=(x+r)^2+z^2$ and $E(M)$ is the complete elliptic integral of the second kind.
Therefore, we perform both the energy and $r$ integrations numerically.

\section{Calculation}\label{calculation}
\begin{figure}[t]
\begin{center}
\includegraphics[scale=0.45]{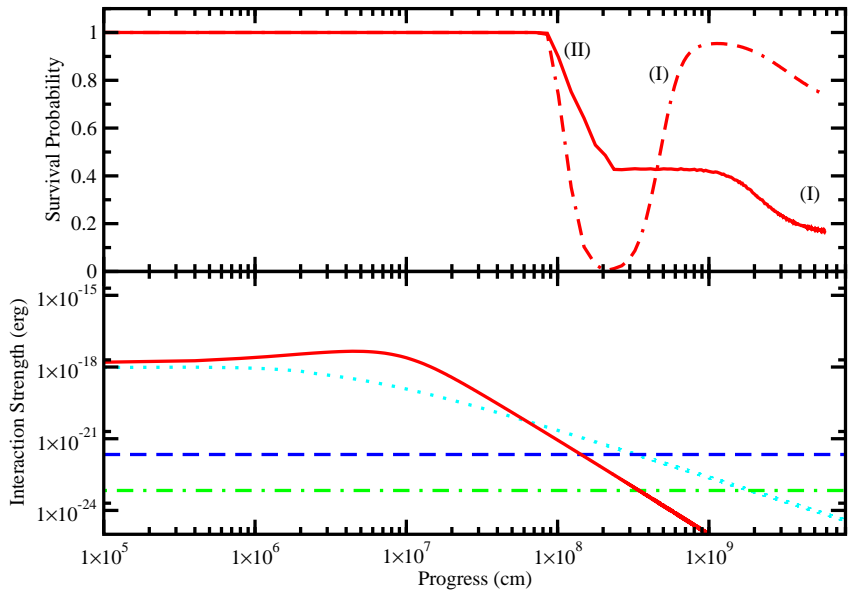}
\caption{{\bf Model A; inverted hierarchy.} Electron neutrino survival probability and strengths of the neutrino-neutrino interaction term for the inverted hierarchy.
Upper plot:  Weighted survival probability of electron neutrinos (red, solid) and electron antineutrinos (red, dot-double-dashed) as a function of the progress along the trajectory.
The weighted survival probabilities are ratios of the energy integrated neutrino and antineutrino capture rates.
All neutrinos began in the electron flavor at $x=z=3\times10^6$ cm. 
Lower plot:  Comparison between the strength of the neutrino-neutrino interaction (red line), the vacuum strengths (blue and green dashed lines), and the neutrino-matter interaction strength (cyan dotted line).
Regions of types (\ref{mswInteractions}) and (\ref{nutations}) are labeled in the upper panel.}
\label{invertedMidSameRnus}
\end{center}
\end{figure}

\begin{figure}[ht]
\begin{center}
\includegraphics[scale=0.45]{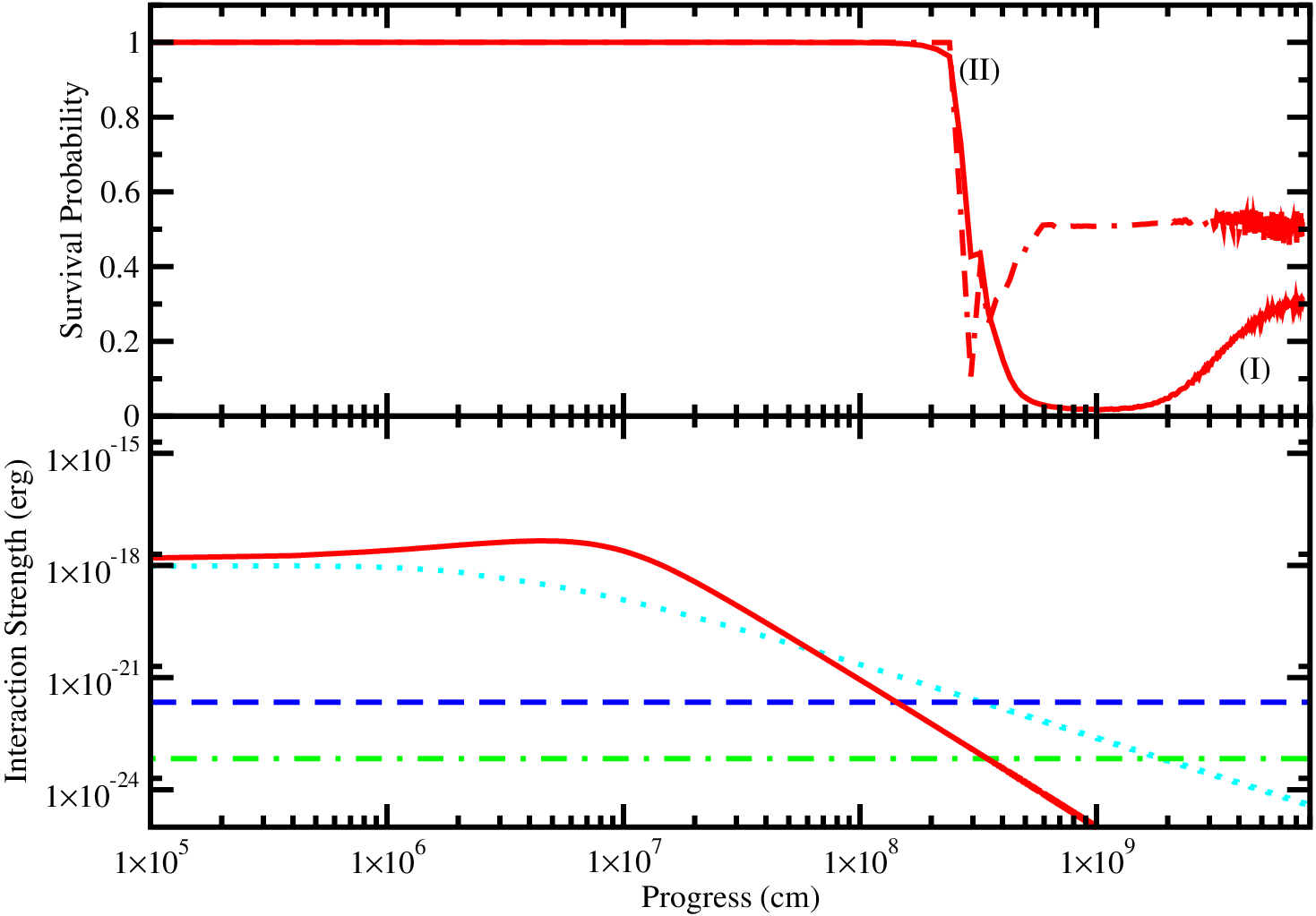}
\caption{{\bf Model A; normal hierarchy.} Weighted electron neutrino survival probability and strengths of the neutrino-neutrino interaction term for the normal hierarchy.
Upper plot: Weighted survival probability of electron neutrinos (red, solid) and electron antineutrinos (red, dot-double-dashed) as a function of the progress along the trajectory.
The weighted survival probabilities are ratios of the energy integrated neutrino and antineutrino capture rates.
All neutrinos began in the electron flavor at $x=z=3\times10^6$ cm. 
Lower plot: Comparison between the strength of the neutrino-neutrino interaction (red line), the vacuum strengths (blue and green dashed lines), and the neutrino-matter interaction strength (cyan dotted line).
Regions of types (\ref{mswInteractions}) and (\ref{nutations}) are labeled in the upper panel.}
\label{normalMidSameRnus}
\end{center}
\end{figure}

\begin{figure}[ht]
\begin{center}
\includegraphics[scale=0.45]{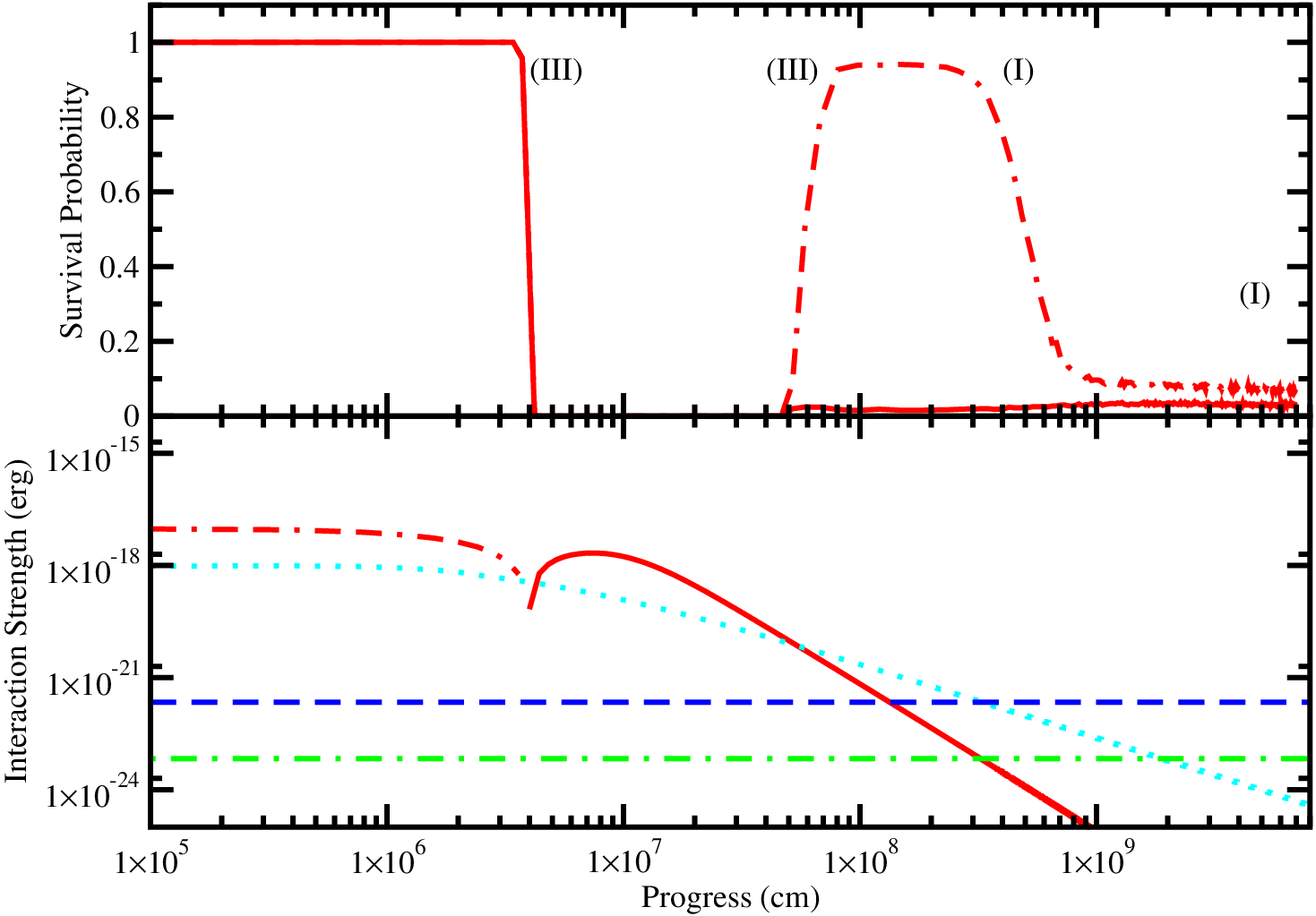}
\caption{{\bf Model B; inverted hierarchy.} Weighted electron neutrino survival probability and strengths of the neutrino-neutrino interaction term the inverted hierarchy.
Upper plot: Weighted survival probability of electron neutrinos  (red, solid) and electron antineutrinos (red, dot-double-dashed) as a function of the progress along the trajectory.
The weighted survival probabilities are ratios of the energy integrated neutrino and antineutrino capture rates. All neutrinos (antineutrinos) began in the electron flavor at $x=z=3\times10^6$ cm. 
Lower plot: Comparison between the strength of the neutrino-neutrino interaction (red line), the vacuum strengths (blue and green dashed lines), and the neutrino-matter interaction strength (cyan dotted line).
Regions of types (\ref{mswInteractions}) and (\ref{cancellationOscillations}) are labeled in the upper panel.}
\label{invertedMidFlop}
\end{center}
\end{figure}

\begin{figure}[ht]
\begin{center}
\includegraphics[scale=0.45]{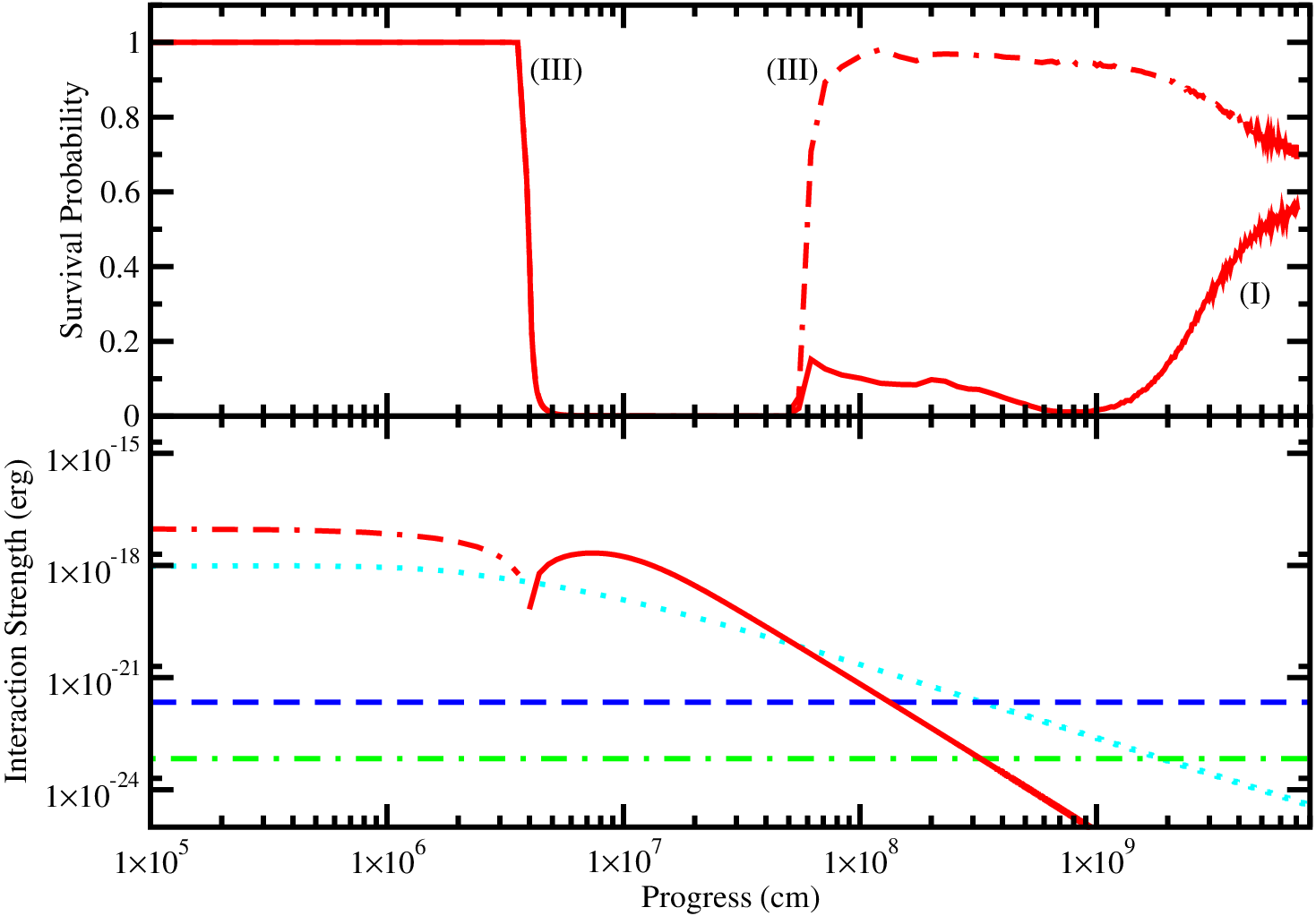}
\caption{{\bf Model B; normal hierarchy.} Weighted electron neutrino survival probability and strengths of the neutrino-neutrino interaction term for the normal hierarchy.
Upper plot: Weighted survival probability of electron neutrinos  (red, solid) and electron antineutrinos (red, dot-double-dashed) as a function of the progress along the trajectory.
The weighted survival probabilities are ratios of the energy integrated neutrino and antineutrino capture rates.
All neutrinos (antineutrinos) began in the electron flavor at $x=z=3\times10^6$ cm. 
Lower plot: Comparison between the strength of the neutrino-neutrino interaction (red line), the vacuum strengths (blue and green dashed lines), and the neutrino-matter interaction strength (cyan dotted line).
Regions of types (\ref{mswInteractions}) and (\ref{cancellationOscillations}) are labeled in the upper panel.}
\label{normalMidFlop}
\end{center}
\end{figure}

Using the disk models described in section \ref{sec:disk-model} and the technique described in section \ref{neutrinos}, we calculate neutrino flavor transformation in both hierarchies. We follow the path of neutrinos from a position of $x=z=3\times10^6$ cm along the $x=z$ trajectory.
We plot the results in Figs. \ref{invertedMidSameRnus}, \ref{normalMidSameRnus}. 
\ref{invertedMidFlop} and \ref{normalMidFlop}.
The upper plots show the neutrino (red, solid) and antineutrino (red, dot-double-dashed) weighted survival probability as a function of progress along the trajectory.
The weighted survival probability is taken to be the ratio between the oscillated and unoscillated neutrino capture rates, i.e. the rates of Eq. \ref{neutrinoCaptureEmission},  and is independent of energy.
The lower plots show the interaction strengths for various contributions to the Hamiltonian in units of ergs.
The red line is $Tr(H_{\nu\nu})$. 
If the term is dominated by neutrinos $Tr(H_{\nu\nu})>0$, the line is solid and if dominated by antineutrinos $Tr(H_{\nu\nu})<0$, it is dot-double-dashed.
The cyan dotted line shows $V_e(t)$ while the blue dashed and green dot-dashed show the vacuum strengths,
\begin{equation}
  \begin{aligned}
    &\frac{\delta m_{13}^2}{2 \left\langle E\right\rangle}, &\frac{\delta m_{12}^2}{2 \left\langle E\right\rangle}
  \end{aligned}
\end{equation}
for the mean energy, $\left\langle E\right\rangle$.

First we use model A where the neutrino disk is larger than the antineutrino disk and slightly cooler, so that the neutrino flux is always larger than the antineutrino flux.
We show the results for each hierarchy in Figs. \ref{invertedMidSameRnus} and \ref{normalMidSameRnus}.
In the inverted hierarchy, dramatic transitions are first seen when electron density drops enough to come close to the $\delta m^2_{13}$ vacuum scale just before $10^8$ cm; in the normal hierarchy, the probability remains almost flat until transitions occur at $\sim 2 \times 10^8$ cm.

Our second model, model B, entails a neutrino disk that is large, but cooler than the antineutrino disk.
The antineutrino disk contributes more to the flux close to its surface but further away, the neutrino disk takes over.
The contribution to the evolution from neutrino-neutrino interaction changes sign somewhere above the disk.
The results are shown in Figs. \ref{invertedMidFlop} and \ref{normalMidFlop}.
The results are similar for both hierarchies. 
The weighted electron neutrino survival probability begins at unity.
As the neutrino-neutrino interaction strength changes sign, the survival probability plummets.
The transition begins when the neutrino-neutrino interaction strength is approximately the same magnitude as the neutrino-matter interaction strength (up to the vacuum contribution) but opposite in sign.
The survival probability rises for the antineutrinos again as neutrino-neutrino interaction strength and the neutrino-matter interaction strength are again the same.

\section{Analysis}\label{analysis}
We see three kinds of regions among our models, and associate them with different circumstances.

\begin{enumerate}[(I)]
  \item Standard MSW transitions that occur when the neutrino-matter interaction strength cancels the vacuum strength. \label{mswInteractions}
  \item Transitions that occur when the neutrino-neutrino interaction strength is about the same scale as the vacuum strength. 
    These are associated with nutation oscillations in the NFIS picture. \label{nutations}
  \item Transitions that occur when the neutrino-neutrino interaction strength is the same size as the neutrino-matter interaction strength, 
   and both are much larger than the vacuum strength.
    These are associated with the canceling of those terms. \label{cancellationOscillations}
\end{enumerate}

Standard MSW transitions, type (\ref{mswInteractions}) are well understood.
They become important to both our models beginning at several times $10^8$ cm in Figs. \ref{invertedMidSameRnus}, \ref{normalMidSameRnus}, 
\ref{invertedMidFlop}, and \ref{normalMidFlop}, when low energy contributions from the neutrino flux begin to hit the MSW resonances.
Neutrinos will undergo resonant transitions when the neutrino-matter interaction strength reaches the vacuum strength, $\delta m_{12}^2/2E$;
The hierarchy determines whether neutrinos or antineutrinos undergo oscillation enhancement at $\delta m_{13}^2/2E$. 
The MSW transitions are energy dependent, but since we have presented survival probabilities with the energy dependence integrated out, 
there is no single position where the effect takes place. 
In Fig. \ref{invertedMidSameRnus} for model A, a type (\ref{mswInteractions}) region associated with $\delta m_{13}^2/2E$ occurs between about $2\times10^8$ cm and $10^9$ cm, 
where the survival probability for electron antineutrinos rises. 
In the same figure, the region associated with $\delta m_{12}^2/2E$ appears between about $10^9$ cm and $10^{10}$ cm 
where the electron neutrino survival probability drops slowly.
Similarly, the type (\ref{mswInteractions}) region associated with $\delta m_{13}^2/2E$ for model B in the inverted hierarchy is where the survival probability
for electron antineutrinos drops between about $2\times10^8$ cm and $10^9$ cm in Fig. \ref{invertedMidFlop}.
In the normal hierarchy, the MSW effects are most important for $\delta m^2_{12}/2E$, again, appearing between about $10^9$ cm and $10^{10}$ cm.
Figure \ref{normalMidSameRnus} shows a small rise in the electron neutrino survival probability in this region for model A,
and Fig. \ref{normalMidFlop} shows a significant rise in the electron neutrino survival probability for model B.

A region of type (\ref{nutations}) appears in our model A in the inverted hierarchy just before $10^8$ 
and in the normal hierarchy somewhat after $10^8$ cm as seen in Figs. \ref{invertedMidSameRnus} and \ref{normalMidSameRnus}.
The transformations occur as the neutrino-neutrino interaction strength approaches the 
vacuum scales, $\delta m_{13}^2/(2\left\langle E\right\rangle)$ and $\delta m_{12}^2/(2\left\langle E\right\rangle)$.
This causes a drop in probability for both neutrinos and antineutrinos in both hierarchies.
Nutation oscillations in an anisotropic environment have been studied previously for the protoneutron star \cite{Hannestad:2006nj, Duan:2007mv}.
In the NFIS picture, as articulated in \cite{Hannestad:2006nj} and \cite{Duan:2007mv}, strong oscillations are expected in such a region and are associated with nutation in the inverted hierarchy, where as smaller oscillations are expected in the normal hierarchy.

Previous calculations for neutrino flavor transformation in disks in the inverted hierarchy \cite{Dasgupta:2008cu} followed along the same lines:  both neutrinos and antineutrinos experienced nearly the same drop in probability in the inverted hierarchy in the type (\ref{nutations}) region.
We verify such a drop, at around $10^8$ cm in Fig. \ref{invertedMidSameRnus} but we find that neutrinos and antineutrinos behave differently.
This is expected since we consider disks with neutrino and antineutrino trapped regions of different sizes (also with black holes at the center).
Thus in our calculations, the effective angles for the neutrinos and antineutrinos are different.

The normal hierarchy has not previously been considered with a disk geometry. However, it has been studied in a spherical geometry, i.e. the protoneutron star.
Bipolar oscillations were studied for the normal hierarchy by Duan, Fuller and Qian \cite{Duan:2005cp}, who found small amplitude oscillations in regions of type (\ref{nutations}) for some parameters.
The amplitude of these oscillations was found to depend on the electron density.
Our density is much closer to the vacuum scale than the suppressive density considered in \cite{Duan:2005cp}.
The smaller densities there were associated with larger oscillations in the normal hierarchy.

In model B, we find little evidence for transitions that correspond with the type (\ref{nutations}) region.

Oscillations arise in a region of type (\ref{cancellationOscillations}) as well.
These occur in the normal and inverted hierarchies in model B, but not in model A.
In model B, the region of type (\ref{cancellationOscillations}) occurs at about $4\times 10^6$ cm and then at $5\times10^7$ cm.
To understand why this type of region can produce oscillations,  we look at two flavors, simplifying the neutrino-neutrino interaction term to show the structure,
\begin{equation}
H_{\nu\nu}(t)=\left(
\begin{array}{cc}
h_{\nu_e}(t)-h_{\bar{\nu}_e}(t) & h_{e\mu}(t) \\
h_{\mu e}(t) &h_{\nu_\mu}(t)-h_{\bar{\nu}_\mu}(t)
\end{array}
\right).\label{simpleHNuNu}
\end{equation}
The contributions from electron neutrinos, muon neutrinos, electron antineutrinos and muon antineutrinos are given by 
$h_{\nu_e}$, $h_{\nu_\mu}$, $h_{\bar{\nu}_e}$, $h_{\bar{\nu}_\mu}$ respectively, so that eg,
\begin{equation}
  h_{\nu_e}(t)=\sqrt{2}G_F\int_0^\infty  dq\int_\Omega \left(1-\cos\Theta_{qq'}\right)\rho_{\nu,ee}(\Omega,q,\mathbf{x},t) dn_\nu(\Omega,q,\mathbf{x},t),
\end{equation}
where we've suppressed the dependence of $h(t)$ on $\mathbf x$.
They correspond to the probability of finding a neutrino or antineutrino in a particular flavor state up to the rest of the interaction scale.
For example, at the start of our trajectory, before any oscillation, there are only electron flavor neutrinos and antineutrinos, 
so that $h_{\nu_\mu}=h_{\bar{\nu}_\mu}=0$ and 
\begin{equation}
  \begin{aligned}
    h_{\nu_e}(t)-h_{\bar{\nu}_e}(t)=\sqrt{2}G_F\int_0^\infty  dq\int_\Omega \left(1-\cos\Theta_{qq'}\right)\left(dn_\nu(\Omega,q,\mathbf{x},t)-dn_{\bar{\nu}}(\Omega,q,\mathbf{x},t)\right).
  \end{aligned}
\end{equation}
At an arbitrary time, the total Hamiltonian will be 
\begin{equation}
\begin{aligned}
  &H_V+H_e(t)+H_{\nu\nu}(t)=&\\
  &\frac{1}{2}\left(
  \begin{array}{cc}
    V_e(t)-\frac{h_{\nu_e}(t)-h_{\nu_\mu}(t)}{2}+\frac{h_{\bar{\nu}_e}(t)-h_{\bar{\nu}_\mu}(t)}{2}-\frac{\delta m^2}{2 E}\cos2\theta_{12} & \frac{ \delta m^2}{2 E}\sin2\theta_{12} +2h_{e\mu}(t)\\
    \frac{\delta m^2}{2 E}\sin2\theta_{12}  + 2h_{\mu e}(t) & \frac{\delta m^2}{2 E}\cos2\theta_{12} -V_e(t)+\frac{h_{\nu_e}(t)-h_{\nu_\mu}(t)}{2}-\frac{h_{\bar{\nu}_e}(t)-h_{\bar{\nu}_\mu}(t)}{2}
  \end{array}
  \right),
\end{aligned}
\end{equation}
where we've subtracted of the trace of the interaction terms which supplies an overall phase.
Like with the MSW resonance, we expect to see large flavor transformation when the diagonal terms vanish, leaving only the flavor mixing off-diagonal terms remaining.
When there are only electron flavor neutrinos and antineutrinos, the interaction term is
\begin{equation}
\begin{aligned}
  &H_V+H_e(t)+H_{\nu\nu}(t)=\\
  &\frac{1}{2}\left(
  \begin{array}{cc}
    V_e(t)+\frac{h_{\nu_e}(t)}{2}-\frac{h_{\bar{\nu}_e}(t)}{2}-\frac{\delta m^2}{2 E}\cos2\theta_{12} & \frac{ \delta m^2}{2 E}\sin2\theta_{12} +2h_{e\mu}(t)\\
    \frac{\delta m^2}{2 E}\sin2\theta_{12}  + 2h_{\mu e}(t) & \frac{\delta m^2}{2 E}\cos2\theta_{12} -V_e(t)-\frac{h_{\nu_e}(t)}{2}+\frac{h_{\bar{\nu}_e}(t)}{2}
  \end{array}
  \right).
\end{aligned}
\end{equation}
Neutrinos and antineutrinos will experience the cancellation slightly differently because of the relative sign of the vacuum portion to the interaction portion.
Nevertheless, since the neutrino-matter interaction strength is very large compared to the vacuum, 
there will be transitions for both when the system is antineutrino dominated, $h_{\bar{\nu}_e}(t)>h_{\nu_e}(t)$. 
These criteria are met in model B at about $4\times10^6$ cm.
We see large flavor transitions for both neutrinos and antineutrinos.
When there are no electron flavor neutrinos, the interaction term becomes
\begin{equation}
\begin{aligned}
  &H_V+H_e(t)+H_{\nu\nu}(t)=\\
  &\frac{1}{2}\left(
  \begin{array}{cc}
    V_e(t)-\frac{h_{\nu_\mu}(t)}{2}+\frac{h_{\bar{\nu}_\mu}(t)}{2}-\frac{\delta m^2}{2 E}\cos2\theta_{12} & \frac{ \delta m^2}{2 E}\sin2\theta_{12} +2h_{e\mu}(t)\\
    \frac{\delta m^2}{2 E}\sin2\theta_{12}  + 2h_{\mu e}(t) & \frac{\delta m^2}{2 E}\cos2\theta_{12} -V_e(t)+\frac{h_{\nu_\mu}(t)}{2}-\frac{h_{\bar{\nu}_\mu}(t)}{2}
  \end{array}
  \right).
\end{aligned}
\end{equation}
In this case, we expect that there will be big transitions in both hierarchies when the system is neutrino dominated, $h_{\nu_\mu}(t)>h_{\bar{\nu}_\mu}(t)$. 
These criteria are met in model B (Figs. \ref{invertedMidFlop} and \ref{normalMidFlop}) 
at about $4\times 10^6$ and $5\times 10^7$ cm and we do indeed see flavor transitions, this time, larger for antineutrinos and smaller for neutrinos.
Note that we do not see large flavor transitions in model A at about $10^6$ cm 
because the neutrinos at that point are all electron flavor and neutrinos outnumber antineutrinos. 
There is no way for the diagonal terms to cancel in model A, until the MSW region at a few times $10^8$ cm. 

In model B, a region of type (\ref{nutations}) occurs during the first type (\ref{cancellationOscillations}) region. 
However, we associated this transition primarily with a type (\ref{cancellationOscillations}) region
since the starting point corresponds to the approximate cancellation of the diagonal term in the Hamiltonian.

\section{Ramifications for Nucleosynthesis}\label{nucleosynthesis}
The neutrinos emerging from the disk will interact with material outflowing from the inner disk, and thus influence any resulting nucleosynthesis.
In this work, we present an example case to illustrate the potential impact.
We begin with a one-dimensional disk model with $\dot{m}=3M_\odot/s$, $a=0$ from \cite{Chen:2006rra}. 
We choose this disk because it has characteristics similar to disks formed from stellar collapse \cite{MacFadyen:1998vz,Popham:1998ab,McLaughlin:2006yy}.
The disk has trapped regions for the electron neutrinos and antineutrinos, but not the mu and tau flavors.
We calculate the neutrino and antineutrino decoupling surfaces as in \cite{Surman:2003qt}.
We then find the single-temperature, flat disk approximations to these surfaces that produces the best match to the full disk neutrino emission.
The resulting parameters---$T_{\nu}=3.2$ MeV, $R_{\nu}=150$ km, $T_{\bar{\nu}}=4.1$ MeV, $R_{\bar{\nu}}=100$ km---are identical to those of model B, 
where antineutrinos dominate near the disk and neutrinos farther out. 

The element synthesis is calculated as in \cite{Surman:2005kf}.
The outflow is taken to be adiabatic and radial, with velocity $v=v_{\infty}(1-r_{0}/r)^{\beta}$, 
where $v_{\infty}$ is the final coasting velocity of $0.1c$, and $\beta$ controls how rapidly the material accelerates.
Close to the disk the the material consists of primarily neutrons and protons, with a ratio set by the forward and reverse weak reactions, Eq. \ref{neutrinoCaptureEmission}. 
The subsequent nucleosynthesis is calculated with the sequence of network codes described in \cite{Surman:2005kf}; neutrino interactions on nucleons are included throughout.

\begin{figure}[ht]
\begin{center}
\includegraphics[scale=0.8]{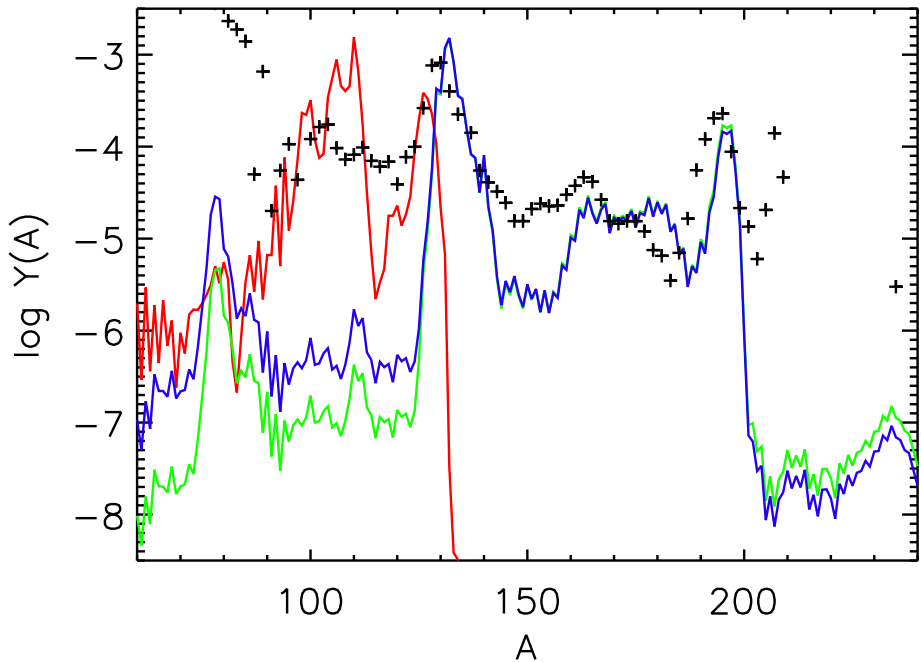}
\caption{Final abundance patterns for simulations with disk outflow parameters $s/k_B=50$ and $\beta=1.4$ in
the absence of neutrino oscillations (red line), with oscillations assuming a normal hierarchy (blue line), and with oscillations
until $r>2\times10^7$ cm, when the neutrino interactions are turned off (green line). 
 The scaled solar $r$-process pattern \cite{Snedan} is included for reference (black crosses).}
\label{abundance}
\end{center}
\end{figure}

\begin{figure}[ht]
\begin{center}
\includegraphics[scale=0.8]{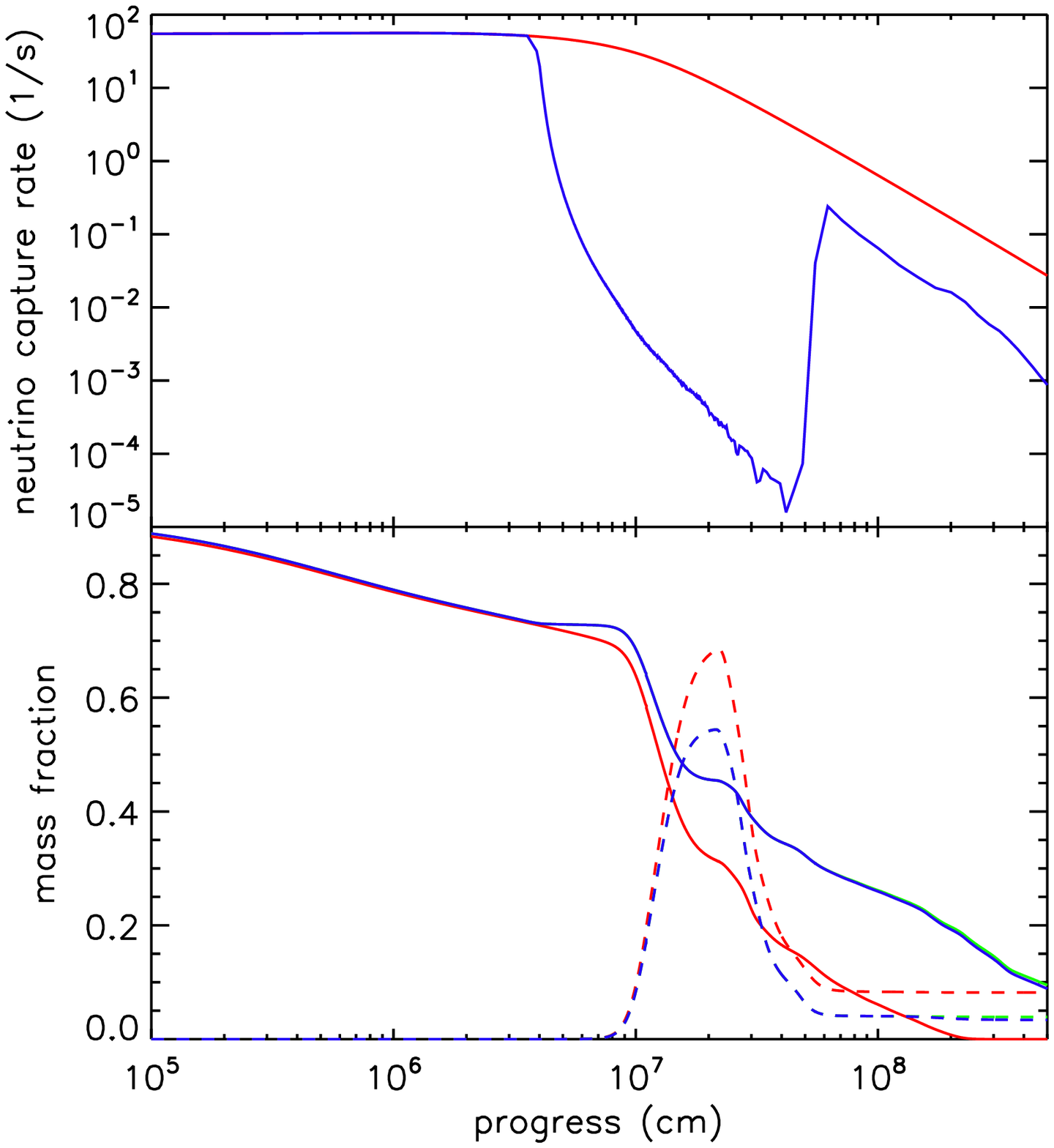}
\caption{The top panel shows the electron neutrino capture rate for the case with no neutrino oscillations (red
line) compared to the case with oscillations in the normal hierarchy (blue line).
The bottom panel shows the mass fractions of
neutrons (solid lines) and alpha particles (dashed lines) for the three nucleosynthesis calculations of Fig. \ref{abundance}}
\label{massFraction}
\end{center}
\end{figure}

If neutrino oscillations are ignored, the outflowing neutron-rich material can synthesize nuclei 
characteristic of the second abundance peak of the $r$-process of nucleosynthesis, as shown in Fig. \ref{abundance} for outflow parameters $s/k_B=50$ and $\beta=1.4$.
The initial neutron-to-proton ratio in the material is greater than one, since the antineutrino capture on protons is initially favored over neutrino capture on neutrons.
All of the protons and most of the neutrons assemble into alphas and heavier `seed' nuclei; 
the $r$-process proceeds when the remaining neutrons are captured on the seeds.
However, once the alphas form, the remaining neutrons are depleted by neutrino interactions on neutrons, 
and additional alphas form immediately from the protons that result.
Thus, there are too many seed nuclei and too few remaining neutrons for a full $r$-process.
The situation changes when neutrino oscillations are taken into account.
The disappearance of neutrinos after they have progressed $4  \times 10^6 $ cm causes the neutrino interaction rate to plummet, as shown in the top panel of Fig. \ref{massFraction}.
Thus, the conversion of neutrons to protons during alpha particle formation is cut off.
Fewer alpha particles and seed nuclei form, and more neutrons remain to capture on these seeds, 
as illustrated in the bottom panel of Fig. \ref{massFraction}.
The result, shown in Fig. \ref{abundance}, is a more robust $r$-process that produces heavy nuclei out to the $A\sim 195$ region.

As the top panel of Fig. \ref{massFraction} shows, the neutrinos reemerge at around 1000 km, 
and thus could potentially slow the $r$-process by converting neutrons to protons at later times in the nucleosynthesis.
To see the extent to which this effect operates, we additionally run the simulation with the neutrinos turned off after the neutrinos have progressed $>2\times10^7$ cm.
The results of this simulation are included in Figs. \ref{abundance} and \ref{massFraction}, 
which show that by this time the capture rates are small and so the impact on the $r$-process is minimal.

\section{Conclusions and Remarks}\label{conclusions}
We have studied models of neutrino and antineutrino emission for accretion disks that encapsulate the qualitative behavior of the neutrino fluxes leaving the disk. 
Two models were examined in detail: those dominated by neutrinos (model A) and those that begin dominated by antineutrinos and end up dominated by neutrinos (model B).
The neutrino dominated disks in the normal hierarchy result in little flavor transition until the neutrino interaction strengths become close to the vacuum strengths.
In both hierarchies they exhibit oscillations in type (\ref{nutations}) (nutation/bipolar) regions.
On the other hand, disks that begin antineutrino dominated and end up neutrino dominated produce large flavor transition 
when the neutrinos flux is about the same as the antineutrino flux.
These transitions are associated with the cancellation of the neutrino and antineutrino terms with the neutrino-matter interaction strength, i.e. a type (\ref{cancellationOscillations}) (matter-neutrino enhanced) region.

The calculations described here can be expanded to more complex scenarios.  We considered disks of
a single temperature, but one should consider also disks with a temperature distribution such
that hotter neutrinos are emitted at the center and cooler neutrinos are emitted at the edges.  The
expected effect would be to shift the interesting transformation behavior nearer to the disk.
We performed our calculations in the single angle approximation, but it would be worthwhile to expand this to multi-angle scenario.  
Based on the arguments in \cite{Duan:2010bg} we expect that single angle calculations will work well as a description of type (\ref{nutations}) transitions just as in the supernova case, although there will be some situations akin to those studied in \cite{Duan:2010bf} when multiangle calculations are necessary.
Type (\ref{cancellationOscillations}) transitions have not been studied from this perspective before.
We compared the position of the type (\ref{cancellationOscillations}) region for the radial neutrino with those coming from various positions on the disk 
and find that this region is at a similar position for all neutrinos.
However, about 30\% of these neutrinos have multiple type (\ref{nutations}) regions.
Thus multi-angle calculations are warranted.
Finally, halo effects have been suggested as a mechanism to alter the simple picture of type \ref{nutations} transformations \cite{Cherry:2012zw}, although no complete calculations exist yet.
Such effects may influence accretion disk
neutrinos as well.

Transitions close to the disk, like those we see in model A, are particularly important because they occur at a time when the neutrinos are influencing nucleosynthesis. 
Using a disk which approximates the type of disk found in a ``collapsar'' scenario \cite{MacFadyen:1998vz}, 
i.e. one that has trapped electron neutrinos and antineutrinos only, with different sized trapping regions, 
we find that the addition of neutrino oscillations enables the formation of $r$-process elements.   These transitions will typically occur close to the disk, where the neutron to proton ratio is being set. The removal of the electron neutrinos as a consequence of this transition, allows the neutron to proton ratio to remain sufficiently high to allow the production of the r-process elements.   This effect should be typical of disks that have type (\ref{cancellationOscillations}) transitions.

\acknowledgements{
We thank North Carolina State University for providing the high performance computational resources necessary for this project. 
This work was supported in part by U.S. DOE Grants No. DE-FG02-02ER41216, DE-SC0004786, DE-FG02-05ER41398, and DE-SC0006417.
}


\end{document}